\documentclass{mn2e}

\usepackage{times}
\usepackage{amssymb}
\usepackage{epsfig}

\begin{document}
\hsize=6truein
          
\title[The UV slopes of high-redshift galaxies]{A critical 
analysis of the UV-continuum slopes  (${\bf \beta}$) of 
high-redshift galaxies;
no evidence (yet) for extreme stellar populations at ${\bf z > 6}$.}

\author[J.S.~Dunlop, et al.]
{J. S. Dunlop$^{1}$\thanks{Email: jsd@roe.ac.uk}, R. J. McLure$^{1}$, 
B. E. Robertson$^2$\thanks{Hubble Fellow}, R. S. Ellis$^2$, D. P. Stark$^3$, \and M. Cirasuolo$^{1,4}$,
L. de Ravel$^1$\\
\footnotesize\\
$^{1}$ SUPA\thanks{Scottish Universities Physics Alliance}, 
Institute for Astronomy, University of Edinburgh, 
Royal Observatory, Edinburgh, EH9 3HJ, UK\\
$^{2}$ Department of Astronomy, California Institute of Technology, Pasadena, CA 91125, USA\\
$^3$Kavli Institute of Cosmology, University of Cambridge, Madingley Road,
Cambridge, CB3 0HA, UK\\
$^4$UK Astronomy Technology Centre, Royal Observatory, Edinburgh, EH9 3HJ}
\maketitle

\begin{abstract}
Following the discovery of the first significant samples of galaxies 
at $z > 6.5$ with Wide Field Camera 3/Infrared (WFC3/IR) 
on board {\it Hubble Space Telescope} ({\it HST}), it has been claimed 
that the faintest high-redshift galaxies display extremely blue ultraviolet (UV)
continuum slopes, with a UV 
power-law index $\beta \simeq -3$ (where $f_{\lambda}~\propto~\lambda^{\beta}$).
Such slopes are bluer than previously reported 
for any other galaxy population, and are most readily explained theoretically by 
extinction-free, young, very low-metallicity stellar populations with a 
high ionizing photon escape fraction. 
Here we undertake a critical study of the evidence 
for such extreme values of $\beta$, combining three new WFC3/IR-selected 
samples of galaxies spanning  nearly two decades 
in UV luminosity over the redshift range
$z \simeq 4.5-8$. We explore the impact of inclusion/exclusion of 
less robust high-redshift candidates, and use the varying depths 
of the samples to explore the effects of noise and selection bias at 
a given ultraviolet luminosity. Simple data-consistency arguments
suggest that artificially blue average values of $\beta$ 
can result when the analysis is extended into the deepest $\simeq 0.5$ mag 
bin of these WFC3/IR-selected galaxy samples, regardless of the actual
luminosity or redshift range probed. By confining attention to robust
high-redshift galaxy candidates, with at least one 8-$\sigma$ detection 
in the WFC3/IR imaging, we find that the average value of $\beta$ is consistent with 
$\langle \beta \rangle = -2.05 \pm 0.10$ over the redshift range 
$z = 5-7$, and the UV absolute magnitude range $-22 < M_{UV,AB} < -18$, and that
$\langle \beta \rangle$ shows no significant trend with either redshift or
$M_{UV}$. 
We create and analyse 
a set of simple end-to-end simulations based on the WFC3/IR+ACS {\it Hubble Ultra Deep Field} (HUDF)
and Early Release Science datasets which 
demonstrate that a bias towards artifically 
low/blue average values of $\beta$ 
is indeed ``expected'' when the UV slope analysis is 
extended towards the source detection threshold, and conclude that there 
is as yet no clear evidence for UV slopes significantly bluer 
than $\beta \simeq -2$, the typical value displayed by the bluest 
star-forming galaxies at more modest redshifts. A robust measurement of
$\langle \beta \rangle$ for the faintest galaxies at $z \simeq 7$ 
(and indeed $z \simeq 8$) 
remains a key observational goal, as it provides 
a fundamental test for high escape fractions from a potentially abundant 
source of reionizing photons. This goal is achievable with {\it HST}, 
but requires still deeper WFC3/IR imaging in the HUDF.

\end{abstract}


\section{INTRODUCTION}

The first galaxies, by definition, are expected to contain very young stellar 
populations of very low metallicity. However, the possibility of detecting 
unambiguous observable signatures of such primordial 
stellar populations with current 
or indeed planned future instrumentation is currently 
a matter of considerable debate.

For example, one long-sought distinctive 
spectral signature of the first generation 
of galaxies is relatively strong HeII emission at $\lambda_{rest} = 1640$\AA\ 
(e.g. Shapley et al. 2003, Nagao et al. 2008, di Serego Alighieri
et al. 2008). However, near-infrared spectroscopy of the sensitivity required 
to detect this line at $z > 7$ will certainly not be available until the 
{\it James Webb Space Telescope} ({\it JWST}), and even then some theoretical
predictions indicate that it is unlikely to be found in detectable objects
(Salvaterra, Ferrara \& Dayal 2011, but see also
Pawlik, Milosavljevic \& Bromm 2011). 

By necessity, therefore, recent attention has focussed on whether 
the broad-band near-infrared photometry which has now been 
successfully used to discover galaxies at $z \simeq 6.5 - 8.5$ 
(e.g. McLure et al. 2010;
Oesch et al. 2010; Bouwens et al. 2010a; Bunker et al. 2010; Finkelstein et al. 2010; 
Vanzella et al. 2011)
can actually be used 
to establish the rest-frame continuum slopes of the highest redshift galaxies.
Specifically, very young, metal-poor stellar populations
are arguably expected to result in substantially bluer continuum slopes around 
$\lambda_{rest} \simeq 1500$\AA\ than have been detected to date in galaxies 
discovered at any lower 
redshift $z < 6.5$ (e.g. Steidel et al. 1999; Meurer et al. 1999; 
Adelberger \& Steidel 2000; Ouchi et al. 2004; Stanway et al. 2005; Bouwens et al. 2006; Hathi et al. 2008; 
Bouwens et al. 2009; Erb et al. 2010).

It has become the normal convention 
to parameterise the ultra-violet continuum slopes of 
galaxies in terms of a power-law index, $\beta$, where $f_{\lambda} \propto
\lambda^{\beta}$ (e.g. Meurer et al. 1999; 
thus, $\beta = -2$ corresponds to a source 
which has a flat spectrum in terms of $f_{\nu}$, and hence 
has zero colour in the AB magnitude system). As discussed by several authors,
while the bluest galaxies observed at $z \simeq 3 - 4$ have $\beta \simeq -2$, 
values as low (i.e. blue) as $\beta = -3$ can in principle be produced 
by a young, low-metallicity stellar population (e.g. 
Bouwens et al. 2010b; Schaerer 2002). However, for 
this idealized prediction to actually be realized in practice, several 
conditions have to be satisfied simultaneously, 
namely i) the stellar population 
has to be very young 
(e.g. $t < 30$\,Myr for $Z \simeq 10^{-3}\,{\rm Z_{\odot}}$, or 
$t < 3$\,Myr for $Z \simeq 10^{-2}\,{\rm Z_{\odot}}$), 
ii) the starlight 
must obviously be completely free from any significant 
dust extinction,
and iii) the starlight must also {\it not} be significantly contaminated by 
(redder) nebular continuum (a condition which has important implications 
for UV photon escape fraction, and hence reionization --  see, for example,
Robertson et al. 2010).

For this reason, the recent report by Bouwens et al. (2010b) (supported to
some extent by Finkelstein et al. 2010) that the 
faintest galaxies detected at $z > 6.5$ do indeed display an average 
value of $\langle \beta \rangle = -3.0 \pm 0.2$ is both exciting and 
arguably surprising enough to merit further detailed and independent 
investigation. This is especially the case because some authors are already 
beginning to assume that the existence of such extreme blue slopes is a robust 
result, already ripe for detailed theoretical interpretation (e.g. Taniguchi et al.
2010).

The aim of this paper is to carefully assess whether the current {\it HST} WFC3
data do indeed provide clear evidence for such extremely 
blue slopes in faint galaxies at $z \simeq 7$. There are a number of potentially
subtle biases which can affect the determination of UV continuum slopes
from the WFC3/IR data, especially when, as is inevitably the case for the faintest
objects, the results have to be based on the {\it average} colours of galaxies
whose individual $\beta$ values have associated errors which can be as large as 
$\Delta \beta \simeq \pm 1.5$. 
To check for, and attempt to quantify, the extent of any such biases we undertake 
two different approaches in this paper. First, we take advantage of the dynamic 
range offered by the available public WFC3/IR imaging to explore how derived 
values of $\beta$ (and average values $\langle \beta \rangle$) depend on 
galaxy candidate robustness and signal:noise ratio as we approach the flux limit 
of a given survey. Second, we undertake and analyse a set of fairly simple (but complete 
end-to-end) simulations 
to explore what apparent values of (and trends in) $\langle \beta \rangle$ 
would be deduced from the existing WFC3/IR data for different assumed 
input values of $\beta = -2, -2.5, -3$ combined with realistic estimates of the 
faint-end slope of the $z \simeq 7$ galaxy luminosity function.

The layout of this paper is as follows.
First, in Section 2 we briefly review how we have selected three new, high-redshift 
galaxy samples from the WFC3/IR+ACS+IRAC imaging of the Hubble Ultra Deep Field (HUDF),
the HUDF Parallel Field 2 (HUDF09-2), and the Early Release Science imaging (ERS) of the 
northern portion of GOODS-South. The reduction of the {\it HST} data, the deconfusion
of the {\it Spitzer} IRAC data, and the extraction, analysis, classification and 
redshift estimation of the galaxies uncovered from this imaging are described in detail in McLure 
et al. (2011), as this 
underpins the extraction of a new robust galaxy sample at 
$6 < z < 8.7$ which is the focus of the McLure et al. (2011) study. In this study we retain 
not only the robust $z > 6$ sources detailed in McLure et al. (2011), but all galaxies from
the larger parent sample with acceptable redshift solutions at $z > 4.5$, which are 
classified as either ROBUST or UNCLEAR. 
This allows us to explore trends in $\beta$ over a 
reasonably wide range in redshift ($5 < z < 8$) and UV luminosity
($-22 < M_{UV,AB} < 18$), 
and also to explore potential biases introduced 
by the exclusion or inclusion of galaxies with less robust photometric redshifts. 
In Section 3 we explain how we determined the rest-frame UV continuum slope,
$\beta$, for the galaxies extracted from the different imaging datasets at 
different redshifts. Then, in Section 4 we present and analyse our results, 
and demonstrate what level of data quality is actually {\it required} to 
achieve internally consistent results between the 
different galaxy samples uncovered from surveys of varying depths. We move on to 
describe and analyse our simulations in Section 5 before discussing the 
implications of our findings in Section 6. A summary of our conclusions is 
then presented in Section 7.
All magnitudes are quoted in the AB system (Oke \& Gunn 1983) and any 
cosmological calculations assume $\Omega_M = 0.3$, $\Omega_{\Lambda} = 0.7$, and 
$H_0 = 70\,{\rm kms^{-1}Mpc^{-1}}$. 

\section{Galaxy Samples}

\begin{figure*}
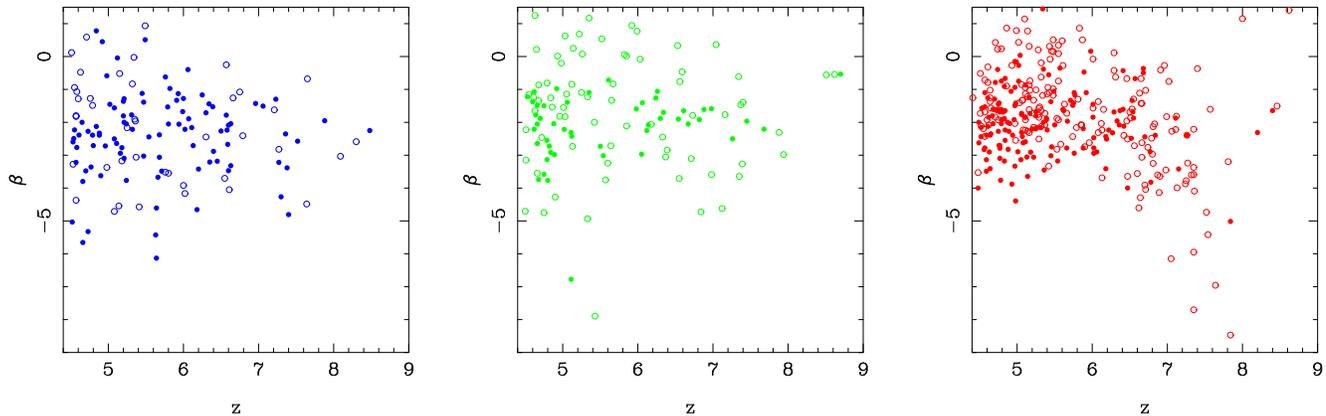

\epsfig{file=fig1a.ps,width=5.4cm,angle=0}
\hspace*{0.5cm}
\epsfig{file=fig1b.ps,width=5.4cm,angle=0}
\hspace*{0.5cm}
\epsfig{file=fig1c.ps,width=5.4cm,angle=0}
\caption{Plots of $\beta$ versus redshift, $z$, 
for all sources in the HUDF sample (left), 
the HUDF09-2 sample (centre), and the ERS sample (right).
Filled symbols
indicate ROBUST sources, open symbols indicate UNCLEAR sources 
which have acceptable alternative low-redshift solutions.}
\end{figure*}

\begin{figure*}
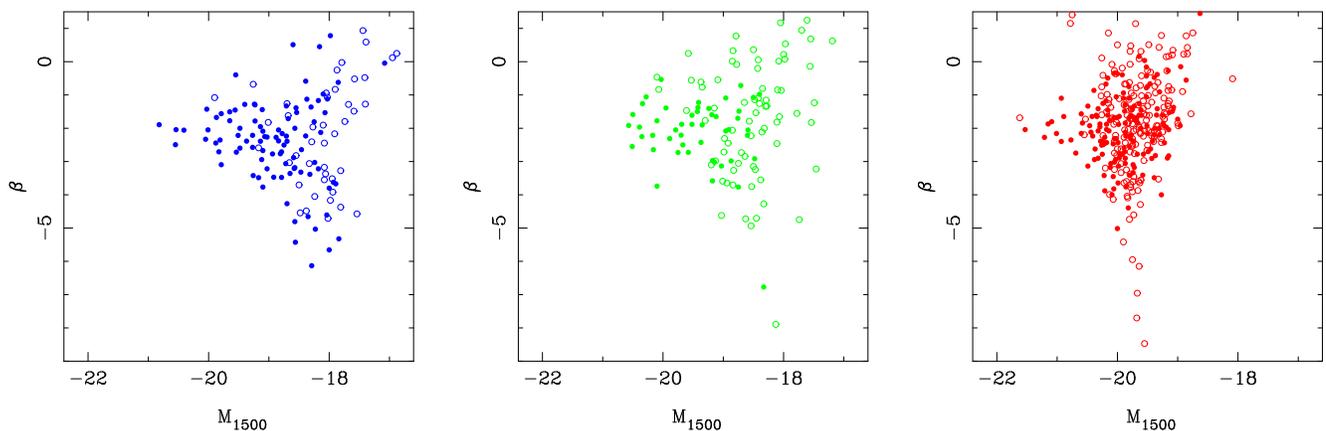

\epsfig{file=fig2a.ps,width=5.4cm,angle=0}
\hspace*{0.5cm}
\epsfig{file=fig2b.ps,width=5.4cm,angle=0}
\hspace*{0.5cm}
\epsfig{file=fig2c.ps,width=5.4cm,angle=0}
\caption{Plots of $\beta$ versus UV absolute magnitude, $M_{1500}$,
for all sources in the HUDF sample (left), 
the HUDF09-2 sample (centre), and the ERS sample (right).
Filled symbols
indicate ROBUST sources, open symbols indicate UNCLEAR sources 
which have acceptable alternative low-redshift solutions.}
\end{figure*}

\subsection{Basic sample production}

The candidate galaxies were all selected from our own reductions 
of the pubicly available near-infrared WFC3/IR imaging of the 
HUDF, ERS and HUDF09-2 fields, as described in McLure et al. (2011)
(we note that the HUDF WFC3/IR imaging is the same year-1, 2009, 
imaging as utilised in McLure et al. 2010).
In brief, candidate selection in all three fields was undertaken by 
first selecting sources with {\sc sextractor} (Bertin \& Arnouts 1996)
down to a deep signal:noise limit in each of 
the WFC3/IR $Y_{105}/Y_{098}$, $J_{125}$ and $H_{160}$ images, 
and then forming the superset of near-infrared selected sources by merging these catalogues.

Then, as again detailed in McLure et al. (2011), 
photometric redshifts (with associated probability
distributions) were derived for all potential sources based on 
0.6-arcsec diameter flux-density measurements made on the available 
{\it HST} ACS optical imaging, the WFC3/IR imaging, and the 
{\it Spitzer} IRAC imaging (after deconfusion of the IRAC images based on the 
WFC3/IR $H_{160}$ or $J_{125}$ data).

The samples were then culled to retain only sources with an acceptable
solution at redshift $z > 4.5$ (i.e. redshift solutions with a formally acceptable
value of $\chi^2$,  typically $\chi^2 < 10$ given the number of data points 
and model free parameters).
All candidate objects were then visually inspected, and rejected from the 
catalogue if they lay too near to the perimeter of the imaging, 
or too close to bright sources
(a cull that is reflected in the effective survey areas quoted by 
McLure et al. 2011). 
Any remaining suspected pseudo sources arising from image artefacts were 
also removed at this stage. 

Finally, all of the ACS+WFC3/IR+IRAC spectral energy distribution
(SED) fits were inspected, 
and the sources classified as either ROBUST or UNCLEAR depending
on whether the alternative low-redshift solution could be excluded 
at $>2$-$\sigma$ on the basis of $\Delta \chi^2 > 4$. We note here that the 
ratio of ROBUST:UNCLEAR sources varies substantially between the fields,
being $\simeq$2:1 in the HUDF, $\simeq$ 1:1 in the ERS, and  $\simeq$1:2 
in HUDF09-2. This is primarily due to the variation in the depths of the 
available optical ACS imaging, relative to the new WFC3/IR near-infrared 
imaging, as discussed further below.

Absolute rest-frame UV magnitudes, $M_{1500}$, have been 
calculated for all objects by integrating the spectral energy distribution
of the best-fitting evolutionary synthesis model (see McLure et al. 2011)
through a synthetic ``narrow-band'' filter of rest-frame width 100\,\AA\, and 
correcting to total magnitude (from the 0.6-arcsec aperture magnitudes 
on which the SED fitting was based) via subtraction of a global aperture 
correction of 0.25\,mag.

\subsection{HUDF}

In the HUDF the high-redshift galaxy sample reported by McLure et al. (2010) has now 
been superceded by the galaxy sample extracted by McLure et al. (2011).
The new parent sample utilised here includes {\it Spitzer} IRAC detections/limits in the 
selection process, and extends to lower redshift to include all objects 
with an acceptable primary redshift solution at $z > 4.5$. 

The resulting HUDF sample contains a total of 147 candidate galaxies with $z_{phot} > 4.5$. Within this master sample, 
95 sources are considered ROBUST according 
to the criterion that the alternative
lower-redshift solution can be rejected with better than 2-$\sigma$ 
confidence (i.e. $\Delta \chi^2 > 4$). The relatively high fraction 
of robust high-redshift sources in this field reflects in large part the 
extreme depth of the asociated optical ACS imaging in the HUDF, which helps 
to establish the robustness of any potential Lyman breaks.

As shown in Figs 1 and 2, the final HUDF galaxy sample at $z > 4.5$
extends to $z > 8$, and  
samples a rest-frame UV luminosity range corresponding 
to $-21 < M_{1500} < -17$ (AB). However,  with one exception, 
ROBUST sources are confined to $-21 < M_{1500} < -18$ (AB).

\subsection{HUDF09-2}

The WFC3/IR imaging of the HUDF09-2 (HUDF Parallel 2) field utilised here
is only 0.07\,mag shallower than the 2009 HUDF WFC3/IR imaging in the $J_{125}$ band,
and covers a similar area. The extracted parent sample is thus comparable in size,
but due to the shallower depth of available optical ACS imaging, the fraction 
of ROBUST:UNCLEAR sources is much lower (see McLure et al. 2011).

The HUDF09-2 sample used here contains 135 candidate galaxies with $z_{phot} > 4.5$.
Within this master sample, 49 sources are considered ROBUST according 
to the criterion that the alternative
lower-redshift solution can be rejected with better than 2-$\sigma$ 
confidence (i.e. $\Delta \chi^2 > 4$).

As shown in Figs 1 and 2, 
the final HUDF09-2 galaxy sample at $z > 4.5$ again extends
to $z > 8$, and samples
a rest-frame UV luminosity range corresponding to $-21 < M_{1500} < -17$ (AB). However, in this case,  
ROBUST sources are basically confined to $-21 < M_{1500} < -18.5$ (AB).

\subsection{ERS}

The ERS WFC3/IR imaging of the northern portion of GOODS-South covers 
an area $\simeq 10$ times larger than each of the above-mentioned ultra-deep
fields, but is typically a magnitude shallower. In addition, the $Y_{098}$ 
filter was utilised in the ERS observations, rather than $Y_{105}$, making the 
$Y$-band imaging even shallower. Because our galaxy selection does 
not involve specific colour cuts, this does not complicate our redshift completeness
(c.f. Oesch et al. 2010, Bouwens et al. 2010a) but this, in combination with optical
data limited to GOODS depth, does mean that about half of the ERS sample 
is classified as UNCLEAR.

The ERS sample used here contains 337 candidate galaxies with $z_{phot} > 4.5$, Within this master sample, 160 sources are considered ROBUST according 
to the criterion that the alternative
lower-redshift solution can be rejected with better than 2-$\sigma$ 
confidence (i.e. $\Delta \chi^2 > 4$).

As shown in Figs 1 and 2, the final ERS galaxy sample at $z > 4.5$ 
extends to $z > 8$, and samples a brighter 
rest-frame UV luminosity range corresponding to 
$-22 < M_{1500} < -19$ (AB).

\begin{figure}
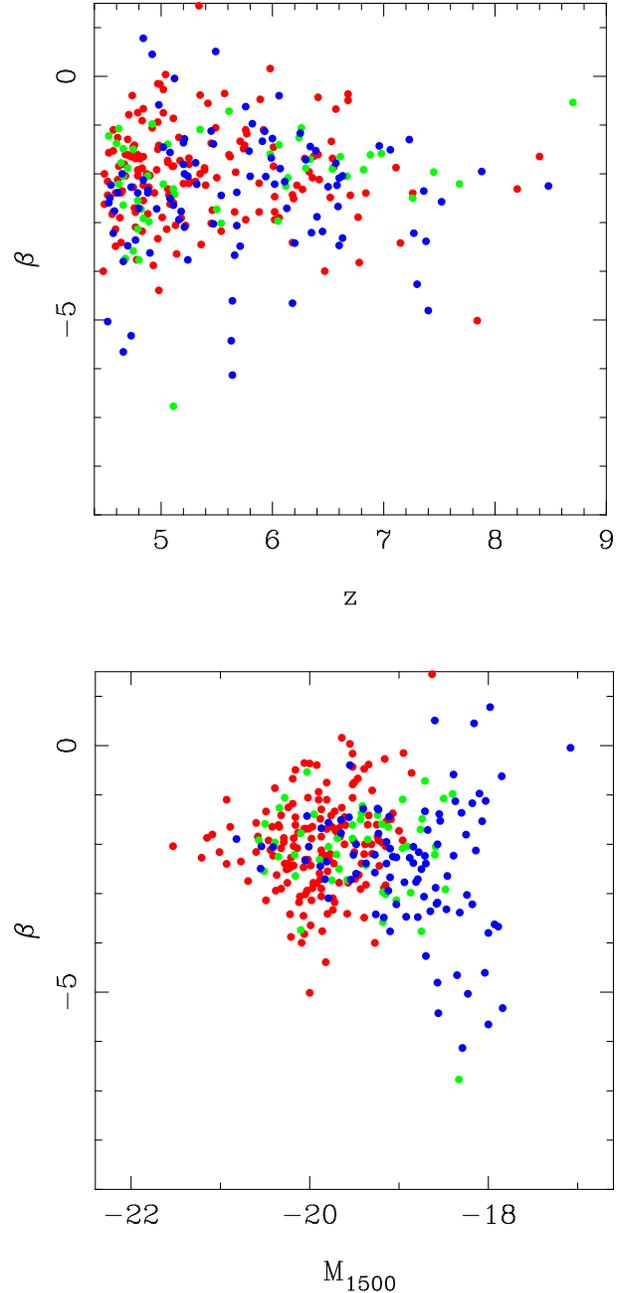

\centerline{\epsfig{file=fig3a.ps,width=8.0cm,angle=0}}
\vspace*{0.8cm}
\centerline{\epsfig{file=fig3b.ps,width=8.0cm,angle=0}}
\caption{Plots of $\beta$ versus redshift $z$, and 
$\beta$ versus UV absolute magnitude $M_{1500}$,
for all the ROBUST sources in the 3 samples
(blue=HUDF, green=HUDF09-2, red=ERS).}
\end{figure}

\begin{figure}
\centerline{\epsfig{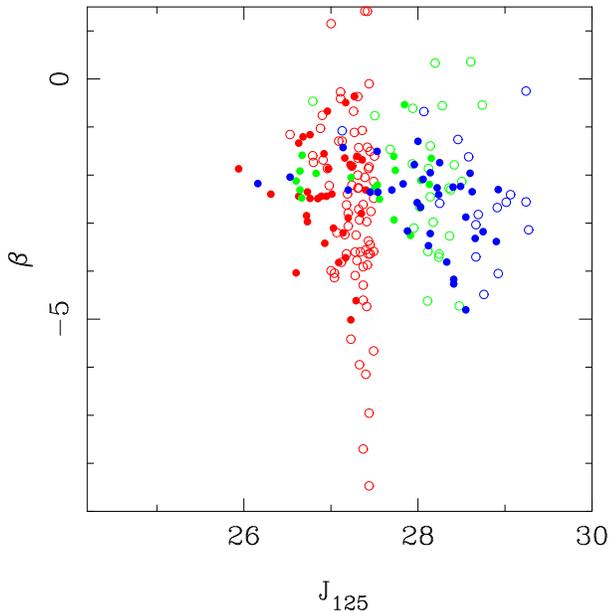}}
\caption{Plot of $\beta$ versus observed $J_{125}$ 
for all sources in the 3 samples with $z_{est} > 6$
(blue=HUDF, green=HUDF09-2, red=ERS). Filled symbols
indicate ROBUST sources while open symbols indicate UNCLEAR sources 
which have acceptable alternative low-redshift solutions.
Apparently extremely blue sources 
with $\beta < -3$ occur at different magnitude ranges 
for the different samples.}
\end{figure}

\section{Measurement of UV continuum slopes}

As already mentioned, 
the standard convention is to characterise the rest-frame UV 
continuum slope via a power-law index, $\beta$, where $f_{\lambda} \propto
\lambda^{\beta}$.

Given the effective wavelengths of the WFC3/IR filters of interest 
here ($Y_{098}$:$\lambda_{eff}$=$9864$\AA; $Y_{105}$:$\lambda_{eff}$=$10552$\AA;
$J_{125}$:$\lambda_{eff}$=$12486$\AA; $H_{160}$:$\lambda_{eff}$=$15369$\AA) 
the relevant conversions from AB mag colours to $\beta$ are:

\begin{equation}
\beta = 4.43(J_{125}-H_{160})-2
\end{equation}
\begin{equation}
\beta = 5.47(Y_{105}-J_{125})-2
\end{equation}
\begin{equation}
\beta = 3.91(Y_{098}-J_{125})-2
\end{equation}

Choosing between the latter two options is dictated by which $Y$-band 
filter was used in the observations, but otherwise the choice is determined
by the estimated redshift of the source. Specifically, at $z_{est} > 6.5$,
both the Lyman-break and any potential Lyman-$\alpha$ emission can enter
the $Y$-band (which cuts in at 9000\AA), thus contaminating any measure of 
$\beta$. Given the uncertainties in $z_{est}$ we therefore use 
equation (1) for any source with $z_{est} > 6.5$,
and equations (2) or (3) as appropriate at lower redshift (to ensure 
we sample comparable rest-frame wavelengths at all redshifts).

We also note that, at $z _{est} > 8$,
both the Lyman-break and any potential Lyman-$\alpha$ emission can enter
the $J_{125}$-band (which cuts in at 11000\AA), and hence, at least for the 
current photometric dataset, any values of $\beta$ 
derived for sources at $z_{est} > 8$ should not be taken seriously
(this is why neither Bouwens et al. (2010b) nor Finkelstein et al. 
(2010) attempted measurement of $\beta$ at $z \simeq 8$). However,
in reality, after application of the galaxy sample quality control 
described below, only one galaxy candidate at $z_{est} > 8$ survives 
(in the HUDF09-2 field) for inclusion in the calculation of average values
of $\beta$.
   Finally, we note that equation (1) above differs very slightly from 
the relation adopted by Bouwens et al. (2010b), which is  $\beta = 4.29(J_{125}-H_{160})-2$,
but the differences in derived values of $\beta$ are completely insignificant
in the current context
(e.g. for $J_{125}-H_{160} = -0.2$, the Bouwens et al. relation yields 
$\beta = -2.86$, while equation (1) yields $\beta = -2.89$).

\section{Results}

\begin{figure*}
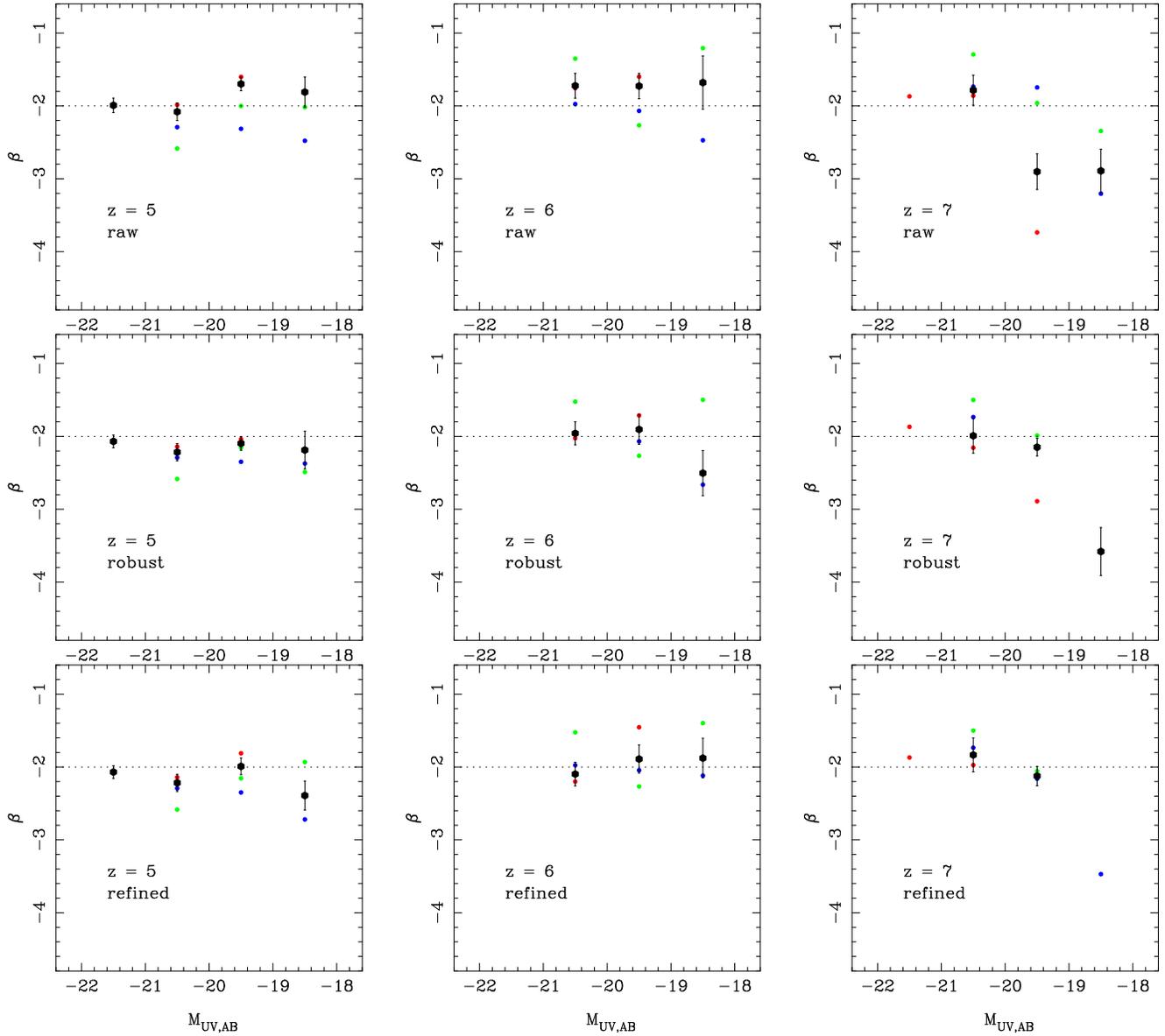

\epsfig{file=fig5a.ps,width=5.4cm,angle=0}
\hspace*{0.5cm}
\epsfig{file=fig5b.ps,width=5.4cm,angle=0}
\hspace*{0.5cm}
\epsfig{file=fig5c.ps,width=5.4cm,angle=0}
\epsfig{file=fig5d.ps,width=5.4cm,angle=0}
\hspace*{0.5cm}
\epsfig{file=fig5e.ps,width=5.4cm,angle=0}
\hspace*{0.5cm}
\epsfig{file=fig5f.ps,width=5.4cm,angle=0}
\epsfig{file=fig5g.ps,width=5.4cm,angle=0}
\hspace*{0.5cm}
\epsfig{file=fig5h.ps,width=5.4cm,angle=0}
\hspace*{0.5cm}
\epsfig{file=fig5i.ps,width=5.4cm,angle=0}
\caption{Plots of average $\langle \beta \rangle$ versus UV absolute magnitude,
at redshift $z \simeq 5$, 6 \& 7 
for all sources (top row),
for ROBUST sources (middle row), and with the additional requirement of at least one 8-$\sigma$ near-infrared 
detection (bottom row). In all panels the small coloured dots 
show the averages derived from each individual field (blue=HUDF, green=HUDF09-2, red=ERS)
while the overall average and standard error are indicated by the black hexagons and 
error bars.}
\end{figure*}

\subsection{Raw results}

In Figs 1 and 2 we plot the raw values of $\beta$ for each source in the 
HUDF, HUDF09-2 and ERS samples versus redshift, $z$, and UV absolute 
magnitude, $M_{1500}$. In each plot the sources classified as ROBUST are 
indicated by the filled symbols, and those classified as UNCLEAR are indicated 
by the open symbols. As well as illustrating the range of redshift and 
UV luminosity probed by each galaxy sample, these plots dramatically illustrate
what an extreme range of apparent individual values of $\beta$ results 
from the photometric uncertainty in colour, especially in the fainter 
luminosity bins probed by each sample. In general it can be seen that a large 
fraction of the faintest galaxies are classified as UNCLEAR. This simply 
reflects the fact that galaxies detected with relatively 
low signal:noise ratio in WFC3/IR, even if
completely undetected at shorter wavelengths, do not display sufficiently
strong breaks in their SEDs to rule out an alternative 
low-redshift solution
(i.e. a Balmer break rather than a Lyman break). However, it is also 
apparent that, even at brighter magnitudes, a large fraction of the
reddest galaxies, with $\beta > -1$, are classified as UNCLEAR. This simply 
reflects a lack of compelling evidence that the continuum above the putative
Lyman break is blue enough 
to rule out a lower-redshift (possibly dusty) solution.
We return to this point later in the paper, when discussing the results of 
the simulations in Sections 5 and 6.

The other point to note from Fig. 2 is that all three samples contain 
ROBUST sources with apparent values of $\beta$ as extreme as $\beta < -5$.
We have not plotted error bars on individual $\beta$ values in 
Fig. 2, but as discussed and plotted 
in McLure et al. (2011), these individual values are of course 
highly uncertain, with $\Delta \beta \simeq \pm 1.5$ or even larger. 
The fact that photometric uncertainty must play 
a major role in producing these extreme values of 
$\beta$ is revealed in Fig. 2 by the fact that 
the plume of extremely blue/low $\beta$ values emerges 
at a different absolute magnitude in the different samples.
 
To illustrate this more clearly we plot all three samples of ROBUST sources
together in Fig. 3. It can be seen that very blue sources with apparent 
values of $\beta < -4$ emerge in the ERS sample at $M_{1500} \simeq -20$, 
whereas in the deeper HUDF data the emergence of such apparently extreme
sources is delayed until $M_{1500} > -18.7$. 
The impact on the scatter in $\beta$ of approaching the flux-density limit 
is perhaps illustrated more clearly in Fig. 4 where 
we confine attention to $z > 6$, and plot $\beta$ versus observed $J_{125}$.

Of course, while photometric scatter inevitably causes the apparent range of 
$\beta$ to rapidly increase as the sample flux limit is approached, there are 
also a large number of galaxies in the final luminosity bin probed by 
each sample, so moderately accurate measures of the average value of $\langle 
\beta \rangle$ can in principle still be derived. Therefore the key issue,
which we now explore, is how the average value of $\langle 
\beta \rangle$ is affected by the inclusion or exclusion of ROBUST/UNCLEAR
sources, and also the signal:noise ratio limit of the galaxy catalogue.
  
\subsection{Quality control}

In Fig. 5 we have divided the three-field sample into 3 redshift bins, and 
binned the data into luminosity bins 1 magnitude wide. We therefore plot
the average value of $\beta$, ($\langle \beta \rangle$) and in each plot 
show the sample-to-sample variation and the final overall average and standard 
error.

The top row of plots shows the results as computed from all the individual
data points shown in Figs. 1 and 2, including both ROBUST and UNCLEAR 
sources. The second row shows the effect of including only the ROBUST sources.
Finally, the third row shows the effect of further limiting the ROBUST samples
to contain only objects which have at least one 8-$\sigma$ near-infrared 
detection in the WFC3/IR data.

This set of plots reveals some interesting trends in the data which can be 
summarized as follows. First, it can be seen that restricting the sample to 
ROBUST sources only has the general effect of moving $\langle \beta \rangle$
to lower (i.e. bluer) 
values at virtually all redshifts and luminosities. Some of this 
effect may be due to the (desirable) removal of lower-redshift interlopers
from the high-redshift galaxy samples. However, some of this shift could 
also be due to a bias introduced by the fact that a blue UV continuum 
slope increases the chance that a source will be classified as ROBUST, 
especially in the absence of a high signal:noise ratio break. This move to the blue 
is most dramatic in the faintest magnitude bin; taken at face value the middle
row of plots in Fig. 5 suggests that galaxies with $M_{UV,AB} \simeq -18.5$
have a typical UV continuum slope which evolves rapidly with redshift,
with $\langle \beta \rangle \simeq -2.2$ at $z \simeq 5$,
$\langle \beta \rangle \simeq -2.5$ at $z \simeq 6$, 
and $\langle \beta \rangle \simeq -3$ at $z \simeq 7$. This is clearly 
fairly similar to the trend reported by Bouwens et al. (2010b).

However, it is also very clear that, especially in the faintest luminosity
bins, there are huge sample-to-sample variations with, for example, 
the ERS sample delivering very blue values of $\langle \beta \rangle$ 
in the $M_{UV,AB} \simeq -19.5$ bin at $z \simeq 7$ compared to either 
of the deeper HUDF and HUDF09-2 samples. This is basically the effect 
of the plume to low values of $\beta$ seen at the ERS flux limit as shown in 
Fig. 4. To reconcile the results from the different samples in the 
luminosity bins in which they overlap, we found it necessary 
to insist on a minimum signal:noise ratio requirement. To avoid introducing 
any further colour bias we simply chose to insist that every object 
retained in the final, refined sample was detected in at least one WFC3/IR 
near-infrared passband at a minimum level of 8-$\sigma$. The impact of 
this further level of quality control is then shown in the bottom row of 
plots in Fig. 5. No longer do the different samples deliver 
substantially different average values of $\langle \beta \rangle$, and it 
can be seen that the very low values of $\langle \beta \rangle$ were indeed
largely resulting from the lowest signal:noise ratio sources. Importantly, with 
this level of further quality control, we are left with only one 
HUDF object in the $M_{UV,AB} \simeq -18.5$ bin at $z \simeq 7$, and hence 
cannot plot a meaningful average value of $\langle \beta \rangle$. At 
$z = 6$, where we can still probe this luminosity bin, the evidence
for $\langle \beta \rangle$ being significantly bluer than $\langle \beta \rangle = 
-2$ has disappeared.

This final result is summarized in Fig. 6, where we overplot the 
dependence of $\langle \beta \rangle$ on $M_{UV,AB}$ as a function of redshift.
The derived datapoints shown in Fig. 6 are tabulated in Table 1.
Clearly, these results are consistent with $\langle \beta \rangle = -2$ 
over the full redshift and luminosity range which can be probed with these
data. At the bright end they are also in good agreement with the results derived by both 
Bouwens et al. (2010b) and Finkelstein et al. (2010), so any disagreement is really 
confined to $M_{UV} > -20$.

Of course, it might be argued that by insisting on rejecting 
the lowest signal:noise ratio sources, we have effectively ``thrown away'' 
the ``evidence'' for how $\beta$ behaves at the faintest luminosities at the
highest redshifts. However, for all the data consistency arguments outlined 
above, we have good reason to suspect that when $\langle \beta \rangle$ 
in a given luminosity bin depends purely on low signal:noise ratio detections, 
its average value may be seriously biased. To 
explore whether this suspicion is fully justified, and to quantify the likely 
magnitude of any such effect, we now describe the creation and analysis of 
a set of simple simulations. 

\section{Simulations}

\subsection{Simulation design}

To explore and attempt to explain the origin of any apparent bias towards 
excessively blue values of $\beta$ as derived from sources extracted with only
$\simeq 4-5$-$\sigma$ photometry, we undertook a set of relatively simple 
simulations. Specifically, we decided to insert into the ERS and 
HUDF {\it HST} (ACS+WFC3) and {\it Spitzer} IRAC images 
a population of galaxies at $z \simeq 7$, with a chosen 
fixed intrinsic 
value of $\beta$, and then reclaimed these sources using exactly the same 
methodology as used to extract and refine the real high-redshift galaxy sample
(i.e., 
including initial use of SExtractor, full ACS-WFC3-IRAC 
photometry from the real images,
derivation of $\chi^2$ versus $z_{phot}$, high-redshift 
sample refinement requiring 
a statistically acceptable solution at $z_{phot} > 6$, and 
final branding as ROBUST or UNCLEAR depending on whether $\Delta \chi^2 > 4$ 
between the low-redshift and high-redshift solutions).

We created three simulations, one in which all galaxies were assigned template
SEDs with $\beta = -2$, one with all galaxies having $\beta = -2.5$, 
and an extreme simulation with $\beta = -3.0$. Perhaps the key 
feature of our simulations is
that we inserted a galaxy population which extended substantially below the 
nominal flux limits of the images, following the form of the McLure et al. 
(2010) $z \simeq 7$ luminosity function down to $J_{125} = 30$(AB).
This is vital to properly simulate the effect of ``flux-boosting'' of 
some subset of the numerous faint sources into the final galaxy sample.
The point here is that, while completely erroneous 5-$\sigma$ sources are 
extremely unlikely, the random flux boosting of, for example, 3-$\sigma$ 
sources to $\simeq 5$-$\sigma$ in {\it either} $J_{125}$ or $H_{160}$ can 
be relatively common when extracting a flux-limited sample 
down to the $\simeq 4 - 5$-$\sigma$ limit in the presence of steep number counts.
Moreover, such flux boosting is highly likely to be accompanied by a 
significant distortion in derived colour, in excess of that 
``expected'' from adding the formal photometry errors in quadrature 
(because it is extremely unlikely that a 
source would fall on, for example, a 2-$\sigma$ positive noise peak in 
both the $J_{125}$ {\it and} $H_{160}$ images). 

\begin{figure}
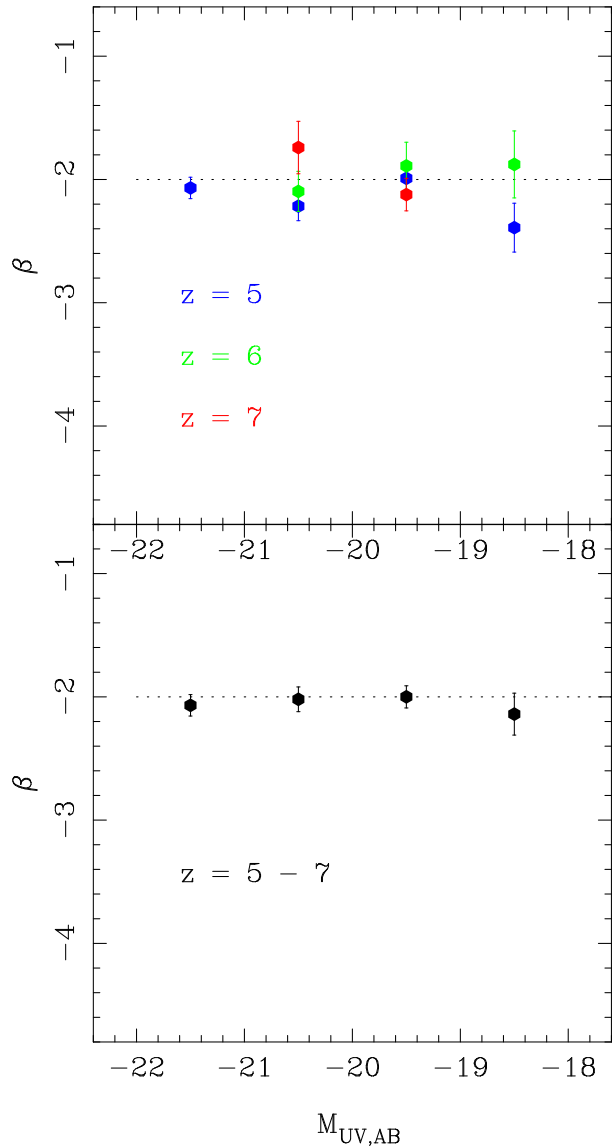

\centerline{\epsfig{file=fig6a.ps,width=8.0cm,angle=0}}
\vspace*{-0.54cm}
\centerline{\epsfig{file=fig6b.ps,width=8.0cm,angle=0}}
\caption{The upper panel shows final average $\langle \beta \rangle$ versus UV absolute magnitude,
$M_{1500}$, at redshift $z \simeq 5$ (blue), 
$z \simeq 6$ (green), and $z \simeq 7$ (red). Values, along with standard errors
in the mean, are plotted 
for any redshift/luminosity bin where the quality control described in the 
text left more than one source. The lower panel 
simply shows the average of the data-points shown in the upper panel,
and thus provides an average value of $\beta$ for each 
luminosity bin, over the redshift range $z = 5 - 7$, including values from 
each redshift bin where this is available.
These data are tabulated in Table 1.
There is no clear evidence 
for any trend with either luminosity or redshift, and all values 
are consistent with $\beta = -2$.}
\end{figure}

\begin{table}
 \begin{center}

\caption{Derived average $\beta$ values and standard errors as a function
of absolute UV magnitude and redshift, as calculated using the robust 
sample of sources with the additional requirement of at least one 8-$\sigma$ near-infrared 
detection. The final column simply gives the average value 
of $\beta$, further averaged over the redshift range $z = 5 - 7$, including values from 
each redshift bin where this is available. The data given here are plotted in Fig.~6.}

\begin{tabular}{lllll}
\hline
$M_{UV}$ &  $\langle \beta \rangle,\,z=5$  &  $\langle \beta \rangle,\, z=6$  &  $\langle \beta \rangle,\,z=7$ &$\langle \langle \beta \rangle \rangle$\\
\hline
$-$21.5  &  $-$2.07$\pm$0.09 &                   &                    & $-$2.07$\pm$0.09   \\
$-$20.5  &  $-$2.22$\pm$0.12 & $-$2.10$\pm$0.16  &   $-$1.74$\pm$0.21 & $-$2.02$\pm$0.10   \\
$-$19.5  &  $-$1.99$\pm$0.11 & $-$1.89$\pm$0.19  &   $-$2.12$\pm$0.13 & $-$2.00$\pm$0.09   \\
$-$18.5  &  $-$2.39$\pm$0.20 & $-$1.88$\pm$0.27  &                    & $-$2.14$\pm$0.16   \\
\hline
\end{tabular}

 \end{center}
\end{table}

\subsection{Simulation results}

We created many realizations for both the ERS and HUDF fields. A typical 
outcome is shown in Fig. 7, for both the $\beta = -2$  and 
$\beta = -2.5$ input catalogues. As in the earlier figures, ERS ``sources''
are indicated in red, HUDF ``sources'' in blue, and ROBUST and UNCLEAR 
high-redshift galaxies are indicated by solid and open symbols respectively.

While comparison with the earlier figures is complicated by the fact that 
Figs 1-3 include galaxies over the redshift range $z = 5-8$, and inevitably 
contain some genuine low-redshift interlopers (at least among the open 
symbols), it can be seen that Fig. 7 reproduces the key features displayed 
by the real data in, for example, Fig. 3. Specifically,
even for input $\beta = -2$,  both the ERS and HUDF simulated 
samples yield galaxies with apparent values of $\beta$ as blue as $\beta 
\simeq -5$ in the faintest luminosity/magnitude bin probed by each sample.
In addition, several of these apparently 
ultra-blue sources are classified as ROBUST.

By contrast, while artificially red sources up to $\beta \simeq 0$ are 
produced by the photometric uncertainties, ultra-red sources are much less 
prevalent,
and red ROBUST sources are very rare (only one ROBUST source in this simulation
is retrieved with $\beta > -1$).

The effect of these distributions of retrieved $\beta$ values on the 
average deduced value of $\langle \beta \rangle$ as a function of UV luminosity
is shown in Fig. 8, again for both the $\beta = -2$ and $\beta = -2.5$ input
scenarios.

The upper panel of Fig. 8 is remarkably similar to the $z \simeq 7$
points plotted in Fig. 1 of Bouwens 
et al. (2010b), and to those given in Fig. 6 of Finkelstein et al. (2010). 
Here the analysis of our $\beta = -2$ simulation has resulted 
in an entirely artificial, apparently monotonic 
luminosity dependence of $\langle \beta \rangle$, 
with $\langle \beta \rangle$ approaching $-3$ in the faintest 
luminosity bin. Only in the brightest bin has the true input value of 
$\beta$ been successfully retrieved. 

It is important to stress that the fact we recover $\langle \beta \rangle 
\simeq -2.4$ 
at $M_{UV} \simeq -19.5$ does not contradict the value of $\langle \beta \rangle \simeq -2.12$ 
we measured from the $z = 7$ data in this bin, as given in Fig. 6 and Table 1. 
As already discussed, to try to minimize bias, these measurements were limited to objects with at least one 
$>8$-$\sigma$ detection in the near-infrared, and even the  $M_{UV} \simeq -19.5$
luminosity bin contains some less significant detections which can bias the result 
to the blue unless filtered out. Thus, our simulation simply implies that, with 
the depth of WFC3/IR data analysed here, unless 
such quality control is applied, a true $\beta \simeq -2$ will result 
in an accurately  measured $\langle \beta \rangle = -2$ 
at $M_{UV} \simeq -20.5$, a somewhat biased
measurement of $\langle \beta \rangle \simeq -2.4$ at $M_{UV} \simeq -19.5$, and a severely biased 
measurement of $\langle \beta \rangle \simeq -3$ at 
$M_{UV} \simeq -18.5$. Thus, our simulation suggests
that the apparent luminosity dependence of $\beta$ with $M_{UV}$ reported by both 
Bouwens et al. (2010b) and Finkelstein et al. (2010) (from the same depth of data) 
is at least {\it not inconsistent} with a true value of $\beta \simeq -2$, independent of 
luminosity.

The lower panel in Fig. 8 simply
shows how even bluer values, with apparent $\langle \beta \rangle < -3$, 
inevitably result when the input value of $\beta$ is  $-2.5$. However, this is clearly 
inconsistent with the data, as the input value of $\beta = -2.5$ is of course 
correctly recovered from the simulation in the brightest luminosity bin, and this is
inconsistent with the observed value of $\beta = -2$ at $M_{UV} = -20.5$.

Interestingly, the retrieved value of $\langle \beta \rangle$ 
in the faintest luminosity bin is not the full 
0.5 lower in the lower panel of Fig. 8 as compared to the upper panel. This implies that 
one cannot easily correct for the bias in a unique way, and that a measured value of  $\langle \beta \rangle \simeq -3$
in this luminosity bin could be consistent with a true $\beta = -2$ or $\beta = -2.5$ within the 
errors. This simply reinforces the need to improve the depth of the WFC3/IR data to enable 
higher signal:noise measurements of $\beta$ in this crucial faint luminosity bin at $z \simeq 7$.
 
\begin{figure}
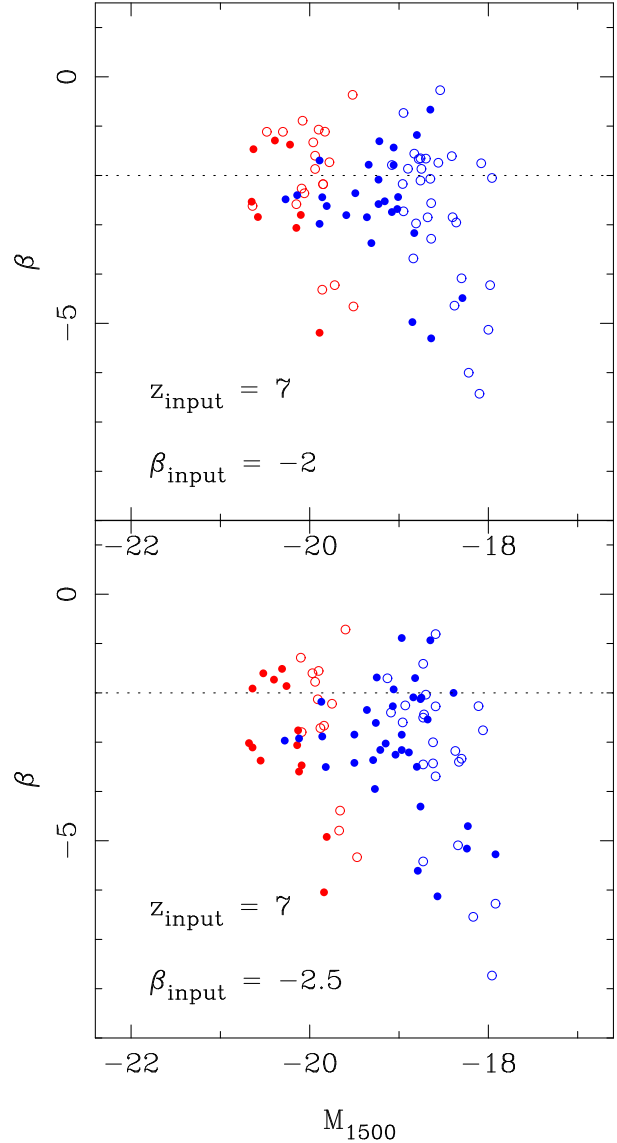

\centerline{\epsfig{file=fig7a.ps,width=8.0cm,angle=0}}
\vspace*{-0.54cm}
\centerline{\epsfig{file=fig7b.ps,width=8.0cm,angle=0}}
\caption{Plots of $\beta$ versus UV absolute magnitude,
$M_{1500}$ as extracted from the ERS (red) and HUDF (blue) WFC3/IR images 
for the $z = 7$ simulated source population described in Section 5. The 
sources plotted in the upper panel were all input with $\beta = -2$, while 
the sources in the lower panel all had $\beta = -2.5$. The simulated 
source population inserted into the real images extended down to an 
input $J_{125}$ magnitude of 30 (AB),
following the form of the $z \simeq 7$ luminosity function derived by McLure
et al. (2010). Sources were then extracted and analysed in exactly the same
way as the real sources; as in Figs 1 and 2, open circles denote UNCLEAR 
sources 
which have acceptable high-redshift solutions, but where the low-redshift 
alternative cannot be formally excluded, while the solid symbols denote 
ROBUST sources in which the alternative lower-redshift solution can be excluded
at $\simeq 2$-$\sigma$.}
\end{figure}

\begin{figure}
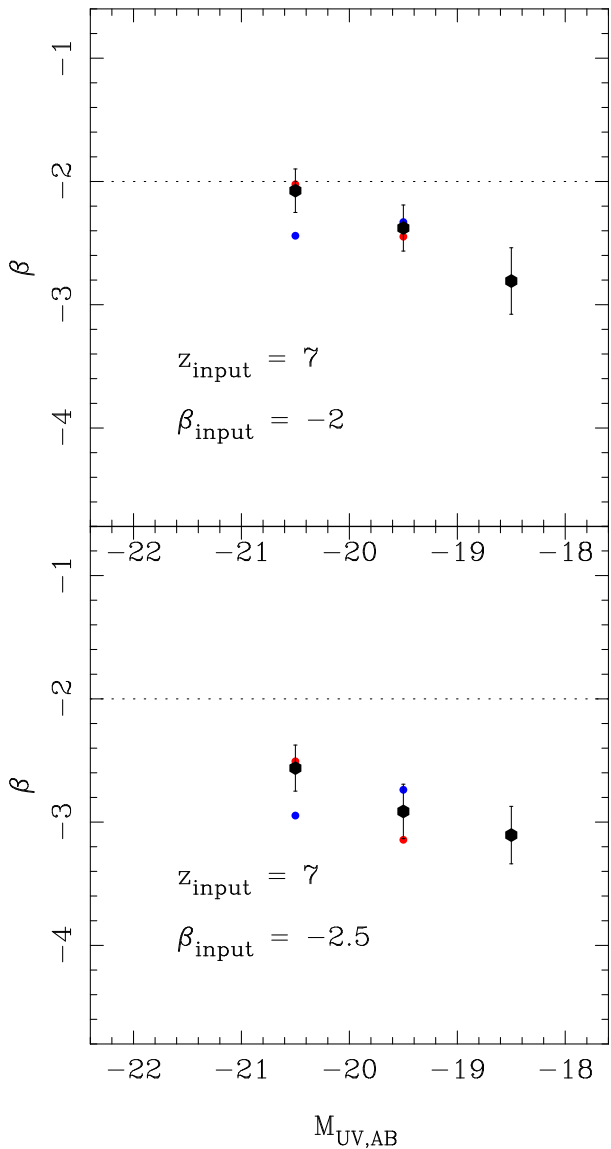

\centerline{\epsfig{file=fig8a.ps,width=8.0cm,angle=0}}
\vspace*{-0.54cm}
\centerline{\epsfig{file=fig8b.ps,width=8.0cm,angle=0}}
\caption{Plots of average $\langle \beta \rangle$ versus UV absolute magnitude,
$M_{1500}$, deduced from the individual $\beta$ values shown in Fig. 9 as 
reclaimed for the two alternative simulated $z \simeq 7$ galaxy populations. 
The upper plot shows how, without careful quality control, a galaxy population
with a true value of $\beta = -2$, independent of $M_{UV}$ yields an apparent
luminosity dependent average value of $\langle \beta \rangle$ which tends to 
$\langle \beta \rangle \simeq -3$ in the faintest luminosity bin from which 
sources can be selected in the current WFC3/IR HUDF data. The similarity between 
this plot and that presented by Bouwens et al. (2010b) (their Fig. 1)
and by Finkelstein et al. (2010) (their Fig. 6) is striking. 
The lower plot shows the same information for the simulated $\beta = -2.5$ population. 
In both plots the correct value of $\langle \beta \rangle$ is only reclaimed 
in the brightest bin, and $\langle \beta \rangle$ becomes progressively 
more biased to the blue with decreasing luminosity.}
\end{figure}

\subsection{The origin of ``$\beta$ bias''}

To explore the origin of the ``$\beta$'' bias so clearly displayed by the 
analysis of our simulations, we take advantage of the fact that 
the ``true'' input UV luminosity of every simulated galaxy is known, and 
explore how derived $\beta$ relates to the level of 
``flux-boosting'' experienced by the simulated sources. 
This is shown in Fig. 9, where the extracted $\beta$ value for all of the 
reclaimed ERS and HUDF high-redshift $\beta = -2$ simulated galaxies is plotted
against  UV luminosity ($\equiv J$-band) flux boost, in magnitudes (here, 
a positive value of ``Boost'' means the reclaimed $J_{125}$ magnitude is 
{\it brighter} than the input value by the plotted magnitude difference).

Both the ERS and HUDF simulated galaxies behave in the same way and show that 
the extremely blue values of $\beta$ almost all result from sources 
which have entered the sample because their ``true'' $J_{125}$ magnitudes have
been boosted by a few tenths of a magnitude (up to $\simeq 0.5$ magnitudes).
This is not really surprising - at the faintest limit a substantial fraction
of the supposedly $5$-$\sigma$ sources in the $J_{125}$ image are 
significantly flux boosted, and random noise dictates that these same 
sources are unlikely to be as extremely flux-boosted at $H_{160}$. Equation 
(1) indicates that a $J_{125}$ flux boost of 0.5 mag, will lead $\beta = -2$ 
to be distorted to $\beta = -4.2$, explaining naturally the behaviour 
displayed in Fig. 9.

However, what is less clear is why comparable flux-boosting in $H_{160}$ has
not produced a comparable population of artificially red objects extending 
to $\beta > 0$. Unless the sample is specifically $J$-band selected 
there is no obvious 
reason why $J_{125}$ flux boosting should be more prevalent 
than $H_{160}$
flux boosting. The answer to this is rather subtle. $H$-band flux boosting 
does occur, and it can be seen that some sources do indeed have their 
$\beta$ values over-estimated, up to values approaching $\beta = 0$. However,
when large/red values of $\beta$ are produced by the noise and flux boosting, 
a significant fraction of the sources start to be classified as 
low-redshift sources by our code, and hence do not appear 
in Fig. 9 (which only contains objects with acceptable reclaimed solutions
at $z_{phot} > 6$). Even among those ``red'' sources 
that do survive, with apparent 
$\beta \simeq -1$, it can be seen that few are classified as robust, simply 
because the measured redder $J-H$ colour permits an acceptable low-redshift 
solution. 

\begin{figure}
\centerline{\epsfig{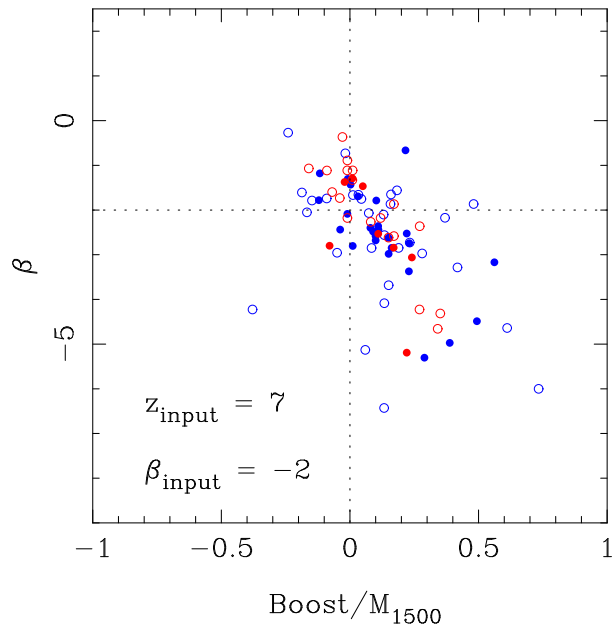}}
\caption{A plot of derived $\beta$ for the simulated 
$z \simeq 7$ $\beta = -2$ sources extracted from the ERS (red) and HUDF (blue) 
WFC3/IR images, versus the UV luminosity ($\equiv J$-band) flux boost, where 
``Boost'' expresses in magnitudes how much brighter the recovered $J$-band 
flux of a given source is as compared to its input flux. Sources with extreme 
apparent values of $\beta$ are largely the result of sources in which 
the true $J$-band flux has been boosted by a few tenths of a magnitude due
to noise in the image (e.g. a 3-$\sigma$ source has been boosted to 5-$\sigma$).
Few sources in which the $H$-band flux has been boosted by comparable amounts 
relative to the $J$-band survive the redshift selection process, and those that
do are generally classified as UNCLEAR (indicated here, as in 
previous plots, by open circles).}
\end{figure}

In summary, as the scatter inevitably rises in the final luminosity bin, 
and is exacerbated by the effects of flux-boosting of fainter sources into 
the (apparently) flux-limited sample, the process of 
high-redshift galaxy selection can clip the red wing 
of the scattered values, and bias the {\it average} $\langle \beta \rangle$
value to the blue. This effect is not really specific to our particular 
method of source selection - a source scattered by $H_{160}$ flux boosting 
to $\beta > 0$ will have an apparent colour $J - H > 0.5$, and is 
therefore less likely to be regarded as a secure high-redshift source 
as selected by standard Lyman-break colour selection techniques. By 
contrast, any artificially ``blue'' galaxy resulting from $J_{125}$ 
flux boosting will almost always be retained, and indeed 
is liable to be classified as ROBUST.
 
We can of course check the extent to which this has happened in our simulated 
galaxy samples. Indeed, for the HUDF simulation shown in Figs 7--9, 
we find that from an input sample of 82 $z = 7$ galaxies, 12 were 
scattered out of the sample to low redshift due, in most cases, to their redder $J-H$ colours.

Finally, for completeness, we show in Fig. 10 the effect of restricting our analysis of the 
$z = 7$,  $\beta = -2$ simulation to sources with at least one $>8$-$\sigma$ detection with WFC3/IR, in effect
replicating our final analysis of the real data as presented in Fig. 6. As with the real data, 
applying this level of quality control leaves us unable to say anything about $\beta$ at $M_{UV} \simeq 
-18.5$, but interestingly (and reassuringly) it also reduces the level of bias in the 
 $M_{UV} \simeq -19.5$ luminosity bin to $\delta \beta \simeq 0.2$. Clearly the results 
presented in Fig. 6 are completely consistent with $\beta = -2$.

\section{Discussion}

Our key results can be summarized as follows.

First we find that, at $z = 5$ and $z = 6$, 
the average value of UV slope is perfectly consistent 
with $\beta = -2$ and displays no 
significant luminosity dependence over the UV luminosity range 
$-22 < M_{1500} < -18$. Second, we find that the same result appears to hold 
at $z \simeq 7$, over the more restricted luminosity range 
$-21 < M_{1500} < -19$, but conclude that no robust statement can yet be made 
about $\langle \beta \rangle$ at fainter luminosities at $z > 6.5$.
Third, we show, both via data consistency arguments from fields of 
varying depth, and from simple (yet realistic, and end-to-end) simulations 
that attempting to extend the measurement of average UV slope into the 
faintest available luminosity bin (as determined by $\simeq 4- 5$-$\sigma$ 
detections) yields values of $\langle \beta \rangle$ 
which are biased to the blue, and can yield apparent average values 
as low as $\langle \beta \rangle \simeq - 3$, even for a true 
input value of $\beta = -2$ for every source.

Thus, while we cannot rule out the recent claims that the faintest galaxies
yet discovered at $z \simeq 7$ have extremely blue slopes, 
$\langle \beta \rangle \simeq -3$, we do show that such extreme values 
of $\langle \beta \rangle$ are {\it not} found 
(from the current data) in 
any luminosity or redshift bin where good-quality photometry 
is available (where ``good'' here means at least one detection in a WFC3/IR 
band at a significance level better than 8-$\sigma$). 
We note here that Finkelstein et al. (2010), while reporting raw results on 
$\langle \beta \rangle$ at $z \simeq 7$ (their Fig. 6) very similar to those 
reported by Bouwens et al. (2010b), derive larger errors on $\langle \beta \rangle$,
and conclude that there is as yet no evidence for a dependence of $\langle \beta \rangle$
on $M_{UV}$ at $z \simeq 7$. 

This, then, provides a very strong and 
clear motivation for still deeper WFC3/IR imaging in the 
HUDF, given the importance of testing the astrophysically important 
possibility that the very faintest high-redshift galaxies do display 
UV slopes significantly bluer that $\langle \beta \rangle = -2.5$, with all the 
associated implications for metallicity, age and ionizing photon escape 
fraction (Bouwens et al. 2010b; Robertson et al. 2010). A depth improvement
of $\simeq 0.5$ mag. would be sufficient to convert most of the current 
$\simeq 5$-$\sigma$ detections to $\simeq 8$-$\sigma$ detections, thus enabling
proper exploration of $\langle \beta \rangle$ down to $M_{UV, AB} \simeq 18$ at
$z \simeq 7$.

Because of its novelty, and potentially crucial implications for reionization, 
we have focussed most of the above discussion, and indeed our simulations, 
on the measurement of $\beta$ at $z \simeq 7$. However, it is also of 
interest to assess how our results at $z \simeq 6$ and $z \simeq 5$ measure
up to previous studies of $\beta$ at these and also lower redshifts.

The most obvious point of comparison is the major study of UV continuum slope
over the redshift range $2 < z < 6$ carried out by Bouwens et al. (2009). 
This work presented extremely good evidence for a luminosity dependence 
of $\beta$ at $z \simeq 2.5$ and $z \simeq 4$. We wish to stress that our 
failure to find any such luminosity dependence in $\beta$ at higher redshifts
should not be taken as casting doubt on these results at lower redshift. In
particular, the evidence presented by Bouwens et al. (2009) for a steady 
decrease in $\langle \beta \rangle$ over the luminosity range 
$-22 < M_{UV} < -17$ appears compelling, based as it is on very large
samples of Lyman-break galaxies in which $\beta$ can be determined purely 
from the {\it optical} photometry (i.e. from {\it HST} ACS $i_{775} - z_{850}$ 
colour). However, this luminosity dependence takes the average value of UV 
slope from $\langle \beta \rangle \simeq -1.3$ 
at $M_{UV} = -22.2$ to $\langle \beta \rangle \simeq -1.9$	
at $M_{UV} = -17.2$, and is attributed by Bouwens et al. (2009) as being 
primarily a result of decreasing dust obscuration with decreasing UV 
luminosity (see also Stark et al. 2010). 
Crucially, even at the faintest luminosities probed at 
$z \simeq 4$, the bluest value of $\langle \beta \rangle$ reported by 
Bouwens et al. (2009) is $\langle \beta \rangle = -2.03 \pm 0.04 \pm 0.15$.

It is by no means obvious that a decrease in dust content with increasing 
redshift should maintain the slope of the $\beta - M_{UV}$ relation, simply
shifting it to more negative values of $\beta$. As already discussed here
and elsewhere, it is relatively straightforward for `normal', essentially 
dust-free stellar populations to produce $\beta = -2$, but the production
of significantly bluer slopes requires different astrophysics in the 
form of very young, very low metallicity stellar populations, with low levels 
of nebular emission.

 \begin{figure}
\centerline{\epsfig{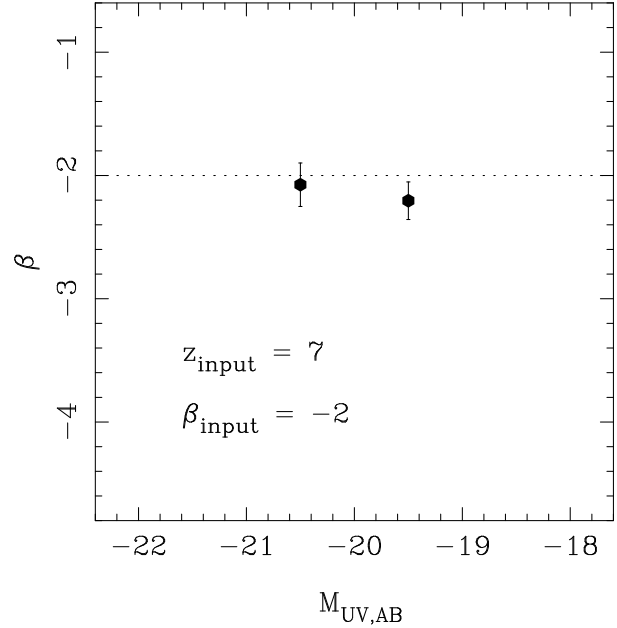}}
\caption{Reclaimed average value of $\langle \beta \rangle$ as a function
of UV luminosity from the $\beta = -2$, $z \simeq 7$ simulation, when attention
is restricted to sources with $> 8$-$\sigma$ detections in $J_{125}$
and/or $H_{160}$. Only a very slight bias to bluer values of $\beta$ remains,
but as with the real galaxy sample at $z \simeq 7$, this restriction 
to decent quality photometry (relatively unaffected by flux boosting)
means that with the current data we can say nothing meaningful about 
the sources with $M_{UV} \simeq -18.5$. To do this requires the current HUDF
imaging to be deepened by a further $\simeq 0.5$ mag. in the relevant WFC3/IR wavelength regime.}
\end{figure}

The results presented by Bouwens et al. (2009) on 
$\beta$ at $z~\simeq~5$ and $z~\simeq~6$ are inevitably 
much more uncertain than those at lower redshift, in part 
because they involved the use of NICMOS data in the measurement of $\beta$.
Nevertheless, at the brighter end of the luminosity range probed, at $M_{UV} 
\simeq -20.5$, our more robust measurements support the conclusion of 
Bouwens et al. (2009) that Lyman break galaxies are bluer at $z \simeq 5-6$ 
than at $z \simeq 4$, with $\langle \beta \rangle$ having moved
from $\langle \beta \rangle = -1.5 \pm 0.15$ at $z \simeq 4$ to 
$\langle \beta \rangle = -2 \pm 0.20$ (see our Fig. 6 and the values
given in Table 4 of Bouwens et al. 2009).

There is thus no serious doubt that the brighter Lyman break galaxies 
have become significantly bluer
with increasing redshift, and the idea that this change is primarily 
due to decreasing dust 
content gains support from the very low (generally negligible) values of 
$A_V$ inferred from the best-fitting SED models at $z > 6.5$ 
deduced by McLure et al. (2011). Therefore the key question now is 
whether the $\beta - M_{UV}$ relation essentially plateaus at 
$\beta \simeq -2$ at $z > 5$ due to the near absence of dust at 
all luminosities, or whether
there is indeed evidence for a continuing dependence of $\beta$ on $M_{UV}$,
albeit perhaps with a different slope. Our own results, as shown in Fig. 6,
support the former scenario, but as already discussed, our analysis also 
emphasizes the vital importance of deeper WFC3/IR data to establish the 
true values of the typical UV slopes of the very faintest galaxies at 
$z \simeq 7$.

\section{Conclusions}

We have undertaken a critical study of the evidence 
for extremely blue UV continuum slopes 
in the highest redshift galaxies, focussing
on the robust determination of the UV power-law index $\beta$ (where $f_{\lambda} \propto \lambda^{\beta}$).
Our analysis is based on three new WFC3/IR-selected 
samples of galaxies spanning  nearly two decades 
in UV luminosity over the redshift range
$z \simeq 4.5-8$ (McLure et al. 2011). We have explored 
the impact of inclusion/exclusion of 
less robust high-redshift candidates, and have used the varying depths 
of the samples to explore the effects of noise and selection bias at 
a given UV luminosity. Simple data-consistency arguments
indicate that artificially blue average values of $\beta$ 
can result when the analysis is extended into the deepest $\simeq 0.5$-magnitude 
bin of these WFC3/IR-selected galaxy samples, regardless of the actual
luminosity or redshift range probed. By confining attention to robust
high-redshift galaxy candidates, with at least one 8-$\sigma$ detection 
in the WFC3/IR imaging, we find that the average value of $\beta$ is consistent with 
$\langle \beta \rangle = -2.05 \pm 0.10$ over the redshift range 
$z = 5-7$, and the UV absolute magnitude range $-22 < M_{UV,AB} < -18$, and that
$\langle \beta \rangle$ shows no significant trend with either redshift or
$M_{UV}$. 

We have created and analysed 
a set of simple end-to-end simulations based on the WFC3/IR+ACS HUDF and ERS datasets 
which 
demonstrate that a bias towards artifically 
low/blue average values of $\beta$ 
is indeed ``expected'' when the UV slope analysis is 
extended towards the source detection threshold, and conclude that there 
is as yet no clear evidence for UV slopes significantly bluer 
than $\beta \simeq -2$, the typical value displayed by the bluest 
star-forming galaxies at more modest redshifts (e.g. NGC1705; $\beta = -2.15$). A robust measurement of
$\langle \beta \rangle$ for the faintest galaxies at $z \simeq 7$ 
(and indeed $z \simeq 8$) 
remains a key observational goal, as it provides 
a fundamental test for high escape fractions from a potentially abundant 
source of reionizing photons. This goal is achievable with {\it HST}, 
but requires still deeper WFC3/IR imaging in the HUDF. We note, however, that, 
due to degeneracies between escape fraction and metallicity, it may prove hard 
to establish robust evidence for a high escape fraction from the measurement of $\beta$ 
{\it unless} extreme values of $\beta = -3$ are indeed confirmed for faint galaxies at $z > 7$ (in which 
case both low metallicity {\it and} a high escape fraction are required).

\section*{ACKNOWLEDGEMENTS}
JSD acknowledges the support of the Royal Society via a Wolfson Research Merit award, and also the support of the European Research Council via the award of an Advanced Grant.
RJM acknowledges the support of the Royal Society via a University Research Fellowship. 
BER is supported by a Hubble Fellowship grant, program number HST-HF-51262.01-A, provided by NASA from the
Space Telescope Science Institute, which is operated by 
which is operated by the Association 
of Universities for Research in Astronomy, Inc, under NASA contract NAS5-26555.
DPS, MC and LdR acknowledge the support of the UK Science \& Technology Facilities Council
via the award of a Post-Doctoral Fellowship, an Advanced Fellowship, and a Post-Doctoral Research Associate position
respectively.
This work is based in part on observations made with the NASA/ESA {\it Hubble Space Telescope}, which is operated by the Association 
of Universities for Research in Astronomy, Inc, under NASA contract NAS5-26555.
This work is also based in part on observations made with the {\it Spitzer Space Telescope}, which is operated by the Jet Propulsion Laboratory, 
California Institute of Technology under NASA contract 1407. 

{}

\end{document}